\documentclass[prl,twocolumn,floatfix,showpacs,superscriptaddress]{revtex4}
\usepackage{amsmath}
\usepackage{graphicx}
\begin{document}

\title{Disorder effects in the quantum kagome antiferromagnet ZnCu$_{3}$(OH)$_6$Cl$_2$ }
\author{M. J. Rozenberg}
\affiliation{Laboratoire de Physique des Solides, CNRS-UMR8502, Universit\'e de Paris-Sud,
Orsay 91405, France.}
\affiliation{Departamento de F\'{\i}sica, FCEN, Universidad de Buenos Aires,
Ciudad Universitaria Pab.I, (1428) Buenos Aires, Argentina.}
\author{R. Chitra}
\affiliation{Laboratoire de Physique Theorique de la Mati\`ere
Condense\'e, UMR 7600, Universite de Pierre et Marie Curie, Jussieu, Paris-75005, France.}

\date{}
\begin{abstract}
Motivated by the recent NMR  experiments on
 ZnCu$_{3}$(OH)$_6$Cl$_2$, 
we study the effect of non-magnetic defects on the antiferromagnetic spin-$\frac12$ kagome lattice.
We use exact diagonalization methods to study the effect of two such defects on
finite size systems. Our results, obtained without adjustable parameters,
are in good quantitative agreement with recent $^{17}$O NMR data. They provide 
support for the experimental interpretation of the presence of defects within the
kagome layers due to Zn/Cu substitutions.
Our results also show that disorder effects become relevant at lower temperatures,
raising questions about the experimental evidence for the absence of an intrinsic spin gap in 
the kagome 2D layers.
 
\end{abstract}
\maketitle

The quest for  spin liquids  in frustrated quantum magnets remains one of the main challenges
in condensed matter physics. Spin  liquids are quantum spin systems where quantum fluctuations 
thwart any kind of ordering of the spins
even in the limit of $T \to 0$.   
Typical criteria for a   candidate system   include: a high degree of frustration, low dimensionality and small spin.
All three features are realized in the spin-$\frac 12$ Heisenberg antiferromagnet on a kagome lattice,
which is a 2D network of corner sharing triangles.
In this context, the recent  discovery  of the compound herbertsmithite, ZnCu$_{3}$(OH)$_6$Cl$_2$
\cite{physicstoday,shores}, presumed to be a perfect physical realization of such a model,
has triggered a spate of experimental studies devoted to
unveiling  the intrinsic behavior of the kagome antiferromagnet.
 A host of experimental techniques, including neutron scattering \cite{helton},
NMR\cite{imai,mendels1} and $\mu$SR\cite{ofer}
have already been used to study the magnetic behavior of this compound.
However, comparisons of experimental and theoretical results are hampered by the fact that
not much is known about the kagome system except that the ground state is a singlet,
the spectrum is characterized by a macroscopic quasi degeneracy of low lying singlet states, a
very short correlation length and possibly a   small gap to
a macroscopic number of quasi degenerate triplet excitations \cite{lhuillier-review,Kagome1}.

Surprisingly, the  plethora of experimental data obtained show that  magnetic degrees of freedom remain active at low frequencies and
temperatures and these do not conform to the naive picture of a gapped system.
This  has  generated a debate  as to whether the experimental systems are pure enough to reveal
intrinsic kagome  behavior, or, alternatively, whether additional interactions that go beyond
simple nearest neighbor antiferromagnetic coupling should be included.
From a chemical perspective, the most delicate aspect is the control
over the positions that  the Cu and Zn  atoms occupy.
The former are magnetic and should be ideally confined to the 2D kagome planes, while the
latter, non-magnetic, should occupy interlayer sites that effectively decouple the planes.
For each exchange of these atoms,  two defects  are created:
(i) the Cu that occupies the Zn site at the interlayer, remains rather weakly coupled and  contributes an effective  ``free'' spin-$\frac 12$ term to the bulk susceptibility,
and (ii) the Zn that occupies a Cu site in a layer, creates  
a non-magnetic defect within the kagome plane which will
likely modify the intrinsic kagome behavior.

On the theoretical front, different ideas were proposed
to account for the unexpected observations: in Ref.~\onlinecite{rrpsingh} it was
shown that
the inclusion of Dyazolinshki-Moriya (DM) interactions may 
enhance the  uniform susceptibility at intermediate temperatures.  On the other
hand, other studies have shown that magnetic defects may also account
for several experimental observations including the uniform susceptibility,
the specific heat \cite{misguich} and the dynamical susceptibility \cite{chitra}.

To elucidate this situation, a rather powerful experimental technique
is the NMR using oxygen isotopes. The oxygen atoms only occupy
sites within the 2D planes and since the hyperfine coupling is very short ranged, oxygen NMR
probes the physics of the plane and is insensitive to the  spin-$\frac 12$
magnetic contributions from misplaced Cu that occupy interlayer sites. Moreover, NMR
probes both static and dynamic behavior.
This decisive experiment was recently performed by Olariu et al. \cite{mendels2}. 
The main results reported were (i) the existence of non-magnetic defects within
the 2D layers, and (ii) the apparent absence of a spin gap.
To comprehend the effect of defects, here we study a spin-$\frac 12$ antiferromagnetic
kagome lattice model with non-magnetic impurities
and obtain the theoretical predictions for the temperature dependence of
the NMR spectra and the nuclear spin relaxation rate $1/T_1$.
The quantitative agreement  between our results  and the experimental
data   unambiguously demonstrates the existence and importance of   non-magnetic defects within the kagome layers
of the herbertsmithite. However,  we also find
that the disorder due to these defects produces a smearing of the NMR line that
is most significant at low temperatures, raising questions about  the experimental
evidence for the
absence of a spin gap in the {\em intrinsic} kagome lattice system.

\begin{figure}
\centerline{\includegraphics[width=0.25\textwidth,angle=0]{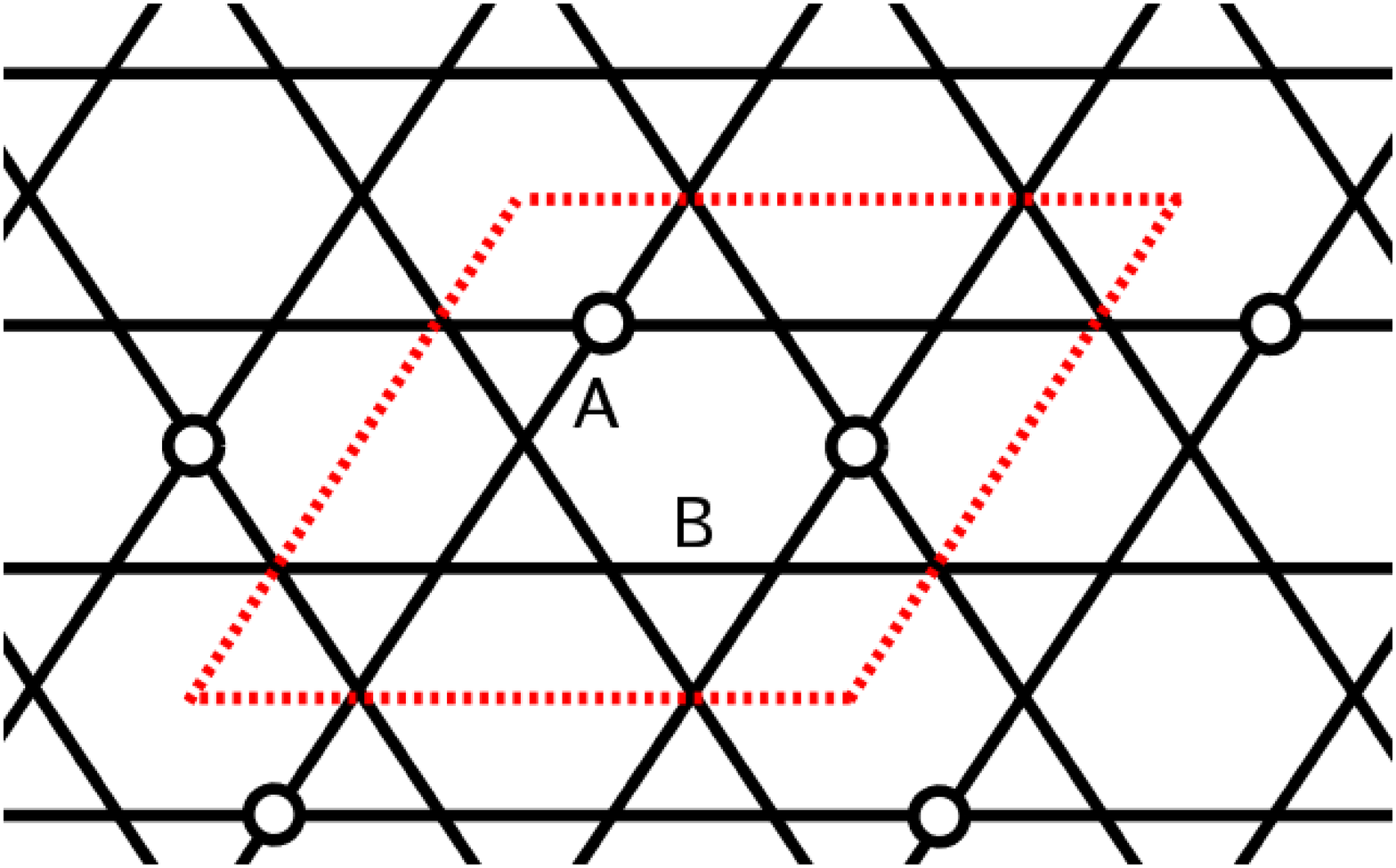}}
\caption{(Color online) Schematic representation of the 12 site kagome lattice
with 2 non-magnetic defects. The bond A denotes a bond with a single spin
and bond B denotes a bond with two spins. For a given configuration of
defects, there are many inequivalent A-bonds and B-bonds.}
\label{fig1}
\end{figure}

The pure hamiltonian is given by the Heisenberg model on the kagome lattice,
\begin{equation}
 {\cal H} = J \sum_{<ij>}  {\bf S}_i \cdot {\bf S}_j
\end{equation}
where, ${\bf S}$ represent the spin-$\frac 12$ operators and $J$ the antiferromagnetic
interactions between two nearest neighbor spins.
Non-magnetic defects are introduced by eliminating a spin operator at a given site.
This model is not very amenable to analytic studies and is usually
studied using numerical methods\cite{lhuillier-review}. 
Here, due to the defects,  we cannot take advantage of the symmetries of the
pure lattice and moreover, as we are interested in the  full temperature dependence,  we  employ  exact diagonalization. We therefore, consider a finite size system of 12 sites as the one schematically
depicted in Fig.~\ref{fig1} with periodic boundary conditions. Note that the 12 site cluster (in the pure case) preserves all
the original symmetries of the full lattice. In addition, since the  magnetic correlation length   in the pure system is  extremely short, of the order of one lattice spacing \cite{leung},
we expect that the 12 site cluster calculation may be adequate for the
specific issues  addressed in the present study \cite{note0}. 

We now study the effect of defects in the 12 site cluster.
We do not consider the case of a single defect, since this would induce local moments which would then contribute  a Curie-like term to the shift.  No such contribution has  been seen in the $^{17}$O NMR study
of Ref.~\onlinecite{mendels2} which probes only the magnetic response in
the kagome layers. In contrast, a free spin-$\frac 12$ contribution was in fact seen in a previous NMR
study\cite{imai}, using Cu and Cl, which probed both in and out of the layer magnetic sites. Thus, the
Curie-like contribution should be attributed to an interlayer Zn/Cu substitution.
In what follows, we consider all possible configurations of two non-magnetic defects. 
Fig.~\ref{fig1} shows an example of one such configuration. 
Two impurities in the twelve site cluster correspond to a defect
concentration of 16\% which is  larger than the experimental estimate of 6-10 \% .  However, due
to the very short correlation length in the pure Kagome system, 
we expect our results to remain valid for a relatively large range of impurity concentrations, 
including the experimental estimate.

The specificity of the $^{17}$O NMR stems from  the structure of ZnCu$_{3}$(OH)$_6$Cl$_2$,
where oxygen is confined to the kagome planes. In a perfectly pure
structure,  each oxygen atom would have two equivalent magnetic Cu(2+) neighbors that are also
located within the 2D kagome layer, and
with whom it interacts via the hyperfine coupling. 
In an NMR experiment, a strong magnetic field is applied producing a local magnetization
of the Cu ions. This magnetization modifies the strength of the hyperfine coupling 
with the nuclear moment, producing a shift in the nuclear magnetic resonance line. Thus, the shift
in the oxygen line is a direct measure of the individual magnetizations, 
and hence of the spin susceptibility, of the neigboring Cu sites
(more precisely of the combined effect of the two neighbors). 
A knowledge of the local fields then gives us information about 
the existence of any underlying spin order.
For example, if two neighboring spin sites were strongly antiferromagnetically correlated 
they would produce no net shift to the central oxygen line.
In a pure system, since all oxygen sites are equivalent,
the NMR spectra should have a single well defined peak with a temperature dependent shift.
However, if non-magnetic defects are present, then the oxygen ions may encounter two types of
magnetic environments: one comprising two nearest-neighbor Cu, and the other with only one  Cu. (The case of two non-magnetic neighbors would produce no NMR shift.)
This would  then be directly reflected in the NMR spectra as a superposition
of two peaks, each one  related to the  respective contribution arising from the 
two kinds of magnetic environments that an oxygen moment may see.

We now present  our  theoretical  predictions for the NMR. 
The only parameter in our model is the magnitude of the antiferromagnetic coupling $J$,
which we set to the experimentally estimated value $\sim 170K$.
The longitudinal magnetic field  in our work $H=0.05J$ is of similar magnitude to the value of 7Tesla  
used in Ref.~\onlinecite{mendels2}. 
To obtain the shifts, we first compute
the finite temperature magnetization at every site $m_i $.
The magnetic shifts $K_{ij}$ at the ``oxygen sites''
(which in the physical structure are
equidistant to two neighboring vertex of the kagome lattice) are  then estimated as the
sum of the two contributions $m_i$ and $m_j$ of the induced moments at the two
sites $i$ and $j$ of each bond of the lattice.  The defect free 2-spin bonds,
such as bond B, depicted in Fig.~\ref{fig1} have in fact two contributions;
however, the 1-spin bonds, such as bond A, have a contribution from 
a sole magnetic moment $m_i$.
Note that  for any given configuration of defects,  since
the translational symmetry
is broken, there are many inequivalent 1-spin
bonds and  2-spin bonds.  The 24 shifts ${K}_{ij}$ for each of the four non-equivalent two-defect 
configurations are computed, and the data at each temperature are condensed into a histogram for
the distribution of the magnitudes of the ${ K}_{ij}$s. 
These histograms can be directly compared to the NMR spectra 
obtained at different temperatures.\cite{note}

\begin{figure}
\centerline{\includegraphics[width=0.4\textwidth,angle=0]{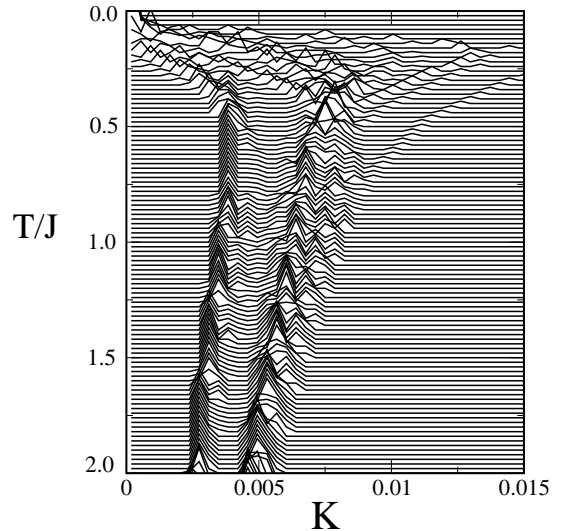}}
\caption{ Histograms for the values of the sum of the 
local magnetization of the spins  on each lattice bond. 
The magnetizations are proportional to the shifts
of NMR. The histograms correspond to the theoretical
estimate for  the NMR spectra. 
}
\label{fig2}
\end{figure}

Fig.~\ref{fig2}  summarizes our results for the predicted NMR spectra
as a function of temperature.
In a pure, defect free system, all sites would be equivalent 
and the histograms would simply show a single sharp peak, whose
position depends on the temperature. Since the shift is proportional
to the local magnetizations due to the external field, its magnitude
is directly proportional to the uniform susceptibility.
However, when defects are introduced, translational invariance is broken
and a distribution of local magnetic moments are induced, which results in the
the histograms of Fig.~\ref{fig2}.
The main features that emerge from the data is the 
presence of two prominent ''ridges'' that can be inferred from the temperature
dependence of the histograms.
The is a clear manifestation of the presence of two types of magnetic environments: some bonds have two
spins (ie, no defect), while others have only one spin (and one defect).  
The position, or shift, of the rightmost ridge is larger by a factor
of two compared to the other one. In fact, the former originates from contributions  of bonds with
two spins while the latter from that of bonds with one spin and one defect.
At high temperatures, as the spins are effectively decoupled, this scaling feature
is intuitively expected. However, the fact that this scaling remains valid well
below  $T/J \sim 1$, is a non-trivial signature of strong frustration and
an extremely short correlation length, 
that  are to be expected in a spin fluid state.
Significantly, this behavior is in very good agreement with the
experimental $^{17}$O NMR data \cite{mendels2}.

\begin{figure}
\centerline{\includegraphics[width=0.4\textwidth,angle=0]{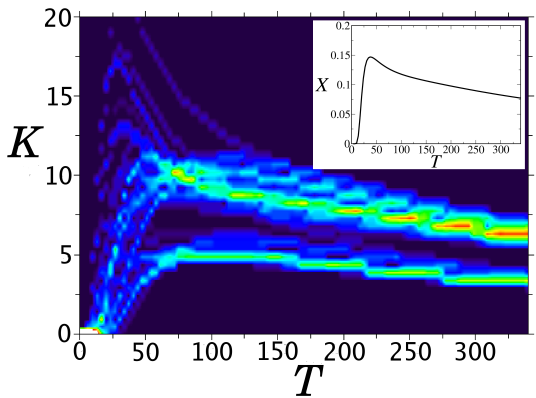}}
\caption{(Color Online) Intensity plot of the histograms of Fig.~\ref{fig2}.
{\cal K} denotes the theoretical estimates for the shifts (in arbitrary units).
The inset shows the uniform magnetic susceptibility $\chi(T)$ for the pure
12 site system.}
\label{fig3}
\end{figure}

In order to visualize the results under a different light, we redisplay the temperature
dependent histograms as an intensity plot in Fig.~\ref{fig3}. For reference,
we also show in the inset the calculated data for the uniform spin susceptibility $\chi(T)$
for the pure system with no defects.  At higher temperatures, $\chi$ shows the Curie tail, while at low $T$ it shows
an exponentially activated behavior, due to the presence of a 
gap in the energy spectrum. 
The main panel shows that the maximum intensity lineshapes  of  the two ridges
roughly follow the behavior of $\chi(T)$. 
However, in contrast to the
 $\chi(T)$ of the pure system, the $T$ dependence of 
both lineshapes  become flat between 70 to 100$K$ where they attain  their 
respective  maximum values. 
Significantly,  a similar flattening in the same temperature regime
around $T \approx 0.5 J$ was  experimentally observed in
Ref.\onlinecite{mendels2}.

Another interesting feature that our results of Figs.~\ref{fig2} and \ref{fig3}
show, is the dramatic smearing of the peaks in the histogram at low temperatures.
Nevertheless, below $T \approx 50K (\approx 0.3J)$ one observes that while the 
lower ridge remains relatively well defined, the
upper one shows a substantial dispersion. This behavior can be qualitatively interpreted as follows:
the lower ridge  arises  from  A bonds which all share the common feature that their
spins are  adjacent to a defect site.  In contrast, the higher  ridge receives contributions from
B bonds whose spins  may be located at different distances from the defects  and hence
 results in a  significant  smearing of the histogram.
 This feature is also qualitatively
observed in the experimental NMR data of Ref.~\onlinecite{mendels2}. There,
the peak of the lower shift (denoted D in Ref.~\onlinecite{mendels2}) remains 
very sharp, while that of the higher shift (denoted M) becomes significantly 
rounded beneath 85$K$.
This defect induced smearing, which is most significant a lower temperatures,
casts doubts on the experimental evidence for the closing of the
{\em intrinsic} spin gap in the kagome system \cite{mendels2}. In fact,
in the measured NMR spectra, the peaks can no longer be resolved beneath 10$K$. 
Our results rather suggest  a scenario  where  the apparent low lying
magnetic excitations reported in the experiment should be attributed to disorder effects stemming
from  uncontrolled substitution of Zn and Cu atoms in the structure of the kagome planes. \cite{note2}

\begin{figure}
\centerline{\includegraphics[width=0.4\textwidth,angle=0]{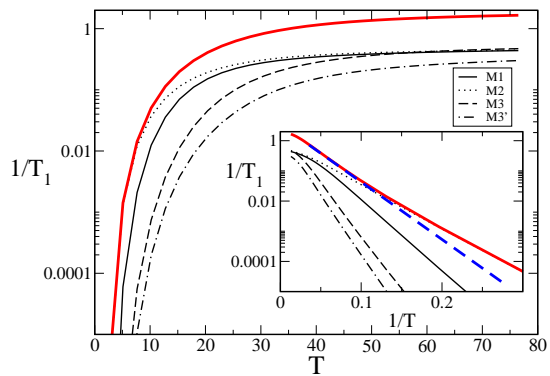}}
\caption{(Color online) The relaxation rate 1/$T_1$  as a function of $T$.
M1, M2, M3 and M3' denote configurations with
the defects at distances 1, 2, 3 and 3 respectively (there are two
inequivalent realizations for distance 3). The solid line is the 
sum of the four contributions. The inset shows the same quantities
on a  semi-log scale as a function of inverse temperature. Straight
lines indicate simple activated behavior. The dashed (blue) straight
line is drawn for reference.
}
\label{fig4}
\end{figure}

Finally, we study the NMR spin relaxation rate estimated as
\begin{equation}
T_1^{-1} \propto  T \frac {{\rm {Im}} \chi(\omega_o)}{\omega_o}
\end{equation}
where $\omega_o$ is the Larmor frequency of the applied external magnetic field and $\chi(\omega)$
is the dynamical  susceptibility.
We compute $\chi(\omega)$ for all 
sites and all defect configurations and evaluate the average value of 1/$T_1$. 
We should note, however, that unlike the magnetization  previously computed for the NMR shifts,  
the  NMR rate probes the low frequency behavior
of the dynamic susceptibility  and  consequently more prone to  finite size effects.
Nonetheless, we expect our results to   provide  useful qualitative insight.
Our calculations for the finite temperature relaxation rate are shown in Fig.~\ref{fig4}.
We plot the data for four particular defect configurations  denoted by the Manhattan distance between the
two defects ( M1, M2, M3 and M3' )
and also the average of the four sets of data (multiplied by 4 for easier visualization).
In the inset of the figure, we show the results as a function of inverse temperature $1/T$  on a semi-log scale. The 
data for M1, M2, M3 and M3' are linear at low $T$ , which is expected 
from simple activated behavior. In contrast, the average data  (solid red line)
clearly deviates from simple activated behavior, displaying a gentle upward
curvature.  These results are  consistent  with the experimental data of Ref.~\onlinecite{mendels2}.
The lack of linearity of the  measured  1/$T_1$ and a putative power law 
fit to the data was interpreted
as a signature of the closing of the {\em intrinsic} spin singlet/triplet gap in the system.
However, our results indicate that an alternative scenario is that
the non-linear behavior is associated with the existence of a multitude of
small activation gaps resulting from the presence of defects.

To conclude, we have studied the effects of non-magnetic
impurity defects on a finite size kagome lattice using exact diagonalization
techniques. With no adjustable parameters our results show a remarkable
agreement with recent $^{17}$O NMR data and  validate
the experimental interpretation  regarding the origin of two prominent
lineshifts in the NMR spectra that were associated with Zn/Cu substitution
in the 2D kagome planes.  
In particular,  the model results  reproduce  the factor of two scaling between the lineshifts which persists down to low temperatures which is indicative of  the spin liquid nature of 
the underlying system.
It also captures the peculiar 
temperature dependence of the measured spin susceptibility and 
  provide insight
on the unequal smearing of the two lineshifts.  On the other hand, our results also cast important doubts on the experimental evidence for
the absence of a spin gap in the system, suggesting that disorder strongly affects
 the low temperature behavior. 
This calls for renewed efforts to achieve a better chemical control on  sample
quality. 


\begin{thebibliography}{10}


\bibitem{physicstoday}
B. Goss Levy, Physics Today, page 16 February 2007.

\bibitem{shores}
M. P. Shores, E. A. Nytko, B. M. Barlett, and D. G. Nocera,
J. Am. Chem. Soc. 127, 13462 (2005).

\bibitem{helton}
J.S. Helton, et al.,
Phys. Rev. Lett. {\bf 98}, 107204 (2007)

\bibitem{ofer}
O. Ofer, et al., cond-mat/0610540.

\bibitem{mendels1}
P. Mendels, F. Bert, M.A. de Vries, A. Olariu, A. Harrison, F. Duc,
J.C. Trombe, J. Lord, A. Amato, C. Baines
Phys. Rev. Lett. {\bf 98}, 077204 (2007)
\bibitem{imai}
T. Imai, E. A. Nytko, B.M. Bartlett, M.P. Shores, D. G. Nocera
ArXiv:cond-mat/0703141.

\bibitem{lhuillier-review}
G. Misguich and C. Lhuillier, ''Frustrated Spin Systems'',
p. 229-306, World Scientific Publishing (2004).
Lhuillier and G. Misguich,
''High magnetic fields'', p. 161-190,
Springer Lecture Notes in Physics (2001).
 
 
\bibitem{Kagome1}
P. Sindzingre, et al. Phys. Rev. Lett. {\bf 84}, 2953 (2000).
P. Lecheminant et al., Phys. Rev. B {\bf 56}, 2521 (1997).
F. Mila, Phys. Rev. Lett. {\bf 81}, 2356 (1998).

\bibitem{rrpsingh}
M. Rigol and R. R. P. Singh. Phys. Rev. Lett. {\bf 98}, 207204 (2007).

\bibitem{misguich}
G. Misguich and P. Sindzingre,  Eur. Phys. J. B {\bf 59}, 305 (2007).




\bibitem{chitra}
R. Chitra and M. J. Rozenberg, Phys. Rev. B {\bf 77}, 052407 (2008).

\bibitem{mendels2}
A. Olariu, P. Mendels, F. Bert, F. Duc, J. C. Trombe, M. A. de Vries, 
and A. Harrison, Phys. Rev. Lett. {\bf 100}, 087202 (2008)


\bibitem{dommange}
S. Dommange, M. Mambrini, B. Normand, and F. Mila,
Phys. Rev. B {\bf 68}, 224416 (2003).

\bibitem{leung} 
P. W. Leung and V. Elser,
Phys. Rev. B {\bf  47}, 5459 (1993).

\bibitem{note0}
Note that similar small  clusters were studied in Ref.\onlinecite{rrpsingh, dommange}.

\bibitem{note}
Note that while in our model the shifts are positive (as  given directly by the local
magnetizations), the experimental shifts happen to be negative due to the physical 
hyperfine coupling constant.

\bibitem{note2}
Although the gap value should be affected by finite size effects, extrapolations from
larger systems data still predict a finite gap. Therefore one may expect our results to be
 qualitatively valid. 
\end{thebibliography}
\end{document}